 \documentclass[twocolumn]{webofc}

\usepackage[varg]{txfonts}   
\usepackage{hyperref}
\usepackage{url}
\hypersetup{colorlinks=true,citecolor=blue,urlcolor=blue,linkcolor=blue}
\begin{document}
\title{Vertex Imaging Hadron Calorimetry Using AI/ML Tools}
%
%

\author{
  \firstname{Nural} \lastname{Akchurin}
  \and
  \firstname{James} \lastname{Cash}
  \and
  \firstname{Jordan} \lastname{Damgov} 
  \and
    \firstname{Xander} \lastname{Delashaw} 
  \and
      \firstname{Kamal} \lastname{Lamichhane} 
  \and
    \firstname{Miles} \lastname{Harris} 
  \and
    \firstname{Mitchell} \lastname{Kelley}
  \and
    \firstname{Shuichi} \lastname{Kunori}
  \fnsep\thanks{\email{shuichi.kunori@ttu.edu}} 
  \and
    \firstname{Harold} \lastname{Mergate-Cacace} 
  \and
    \firstname{Timo} \lastname{Peltola} 
  \and
    \firstname{Odin} \lastname{Schneider} 
  \and
  \firstname{Julian} \lastname{Sewell}
}

\institute{Advanced Particle Detector Laboratory, Department of Physics and Astronomy, Texas Tech University, Lubbock, TX, USA}

\abstract{The fluctuations in energy loss to processes that do not generate measurable signals, such as binding energy losses, set the limit on achievable hadronic energy resolution in traditional energy reconstruction techniques. The correlation between the number of hadronic interaction vertices in a shower and the invisible energy is found to be strong and is used to estimate the invisible energy fraction in highly granular calorimeters in short time intervals (<10 ns). We simulated images of hadronic showers using GEANT4 and deployed a neural network to analyze the images for energy regression. The neural network-based approach results in a significant improvement in energy resolution, from 13\% to 4\% in the case of a Cherenkov calorimeter for 100 GeV pion showers. We discuss the significance of the phenomena responsible for this improvement and the plans for experimental verification of these results and further development.
}
\maketitle
\section{Introduction}
\label{intro1}
Main goal of hadron calorimeter is to measure the high-energy quarks and gluons produced in particle collisions in experiments. Each quark or gluon appears in the detector as a stream of hadrons, called the \textit{ jet}, which consists primarily of photons from $\pi^0$ decays and other hadrons $\pi^{\pm},K^{\pm,0},n,p$. The former forms the electromagnetic shower, and the latter forms the hadronic shower in a calorimeter. The calorimeter response to a photon or a hadron tends to be different, typically (photon response)$>$(hadron response). We discuss the role of ``invisible energy'' in response differences in the next sections.  Two methods have been introduced to overcome the problem of invisible energy. The \textit{Compensation} method was introduced in the late 1970s and implemented the average energy compensation in hardware. The \textit{Dual Readout (DR)} method was introduced in the early 2000s and applies event-by-event compensation using a scintillation signal and a Cherenkov signal in offline software. 

We started the CaloX project at the Advanced Particle Detector Laboratory at TTU in 2019. The goal was to develop a novel calorimetry concept that utilizes new technologies such as AI/ML techniques, high granularity of the calorimeter, fast photo detectors, and high-density electronics. We aimed for a signal collection time of less than 10 ns, which is well below the collision time frequency at the LHC (40 MHz, 25 ns) and the future collider FCC-hh. We reported our earlier results in \cite{RefJ01,RefJ02}.  In the following sections, we discuss the sources of invisible energy and our concept of fast calorimetry for future collider experiments.

\section{Vertex Imaging Calorimetry}
\label{vtx}
\subsection{Invisible Energy at Vertex}
\begin{figure}
\centering
\includegraphics[width=8cm,clip]{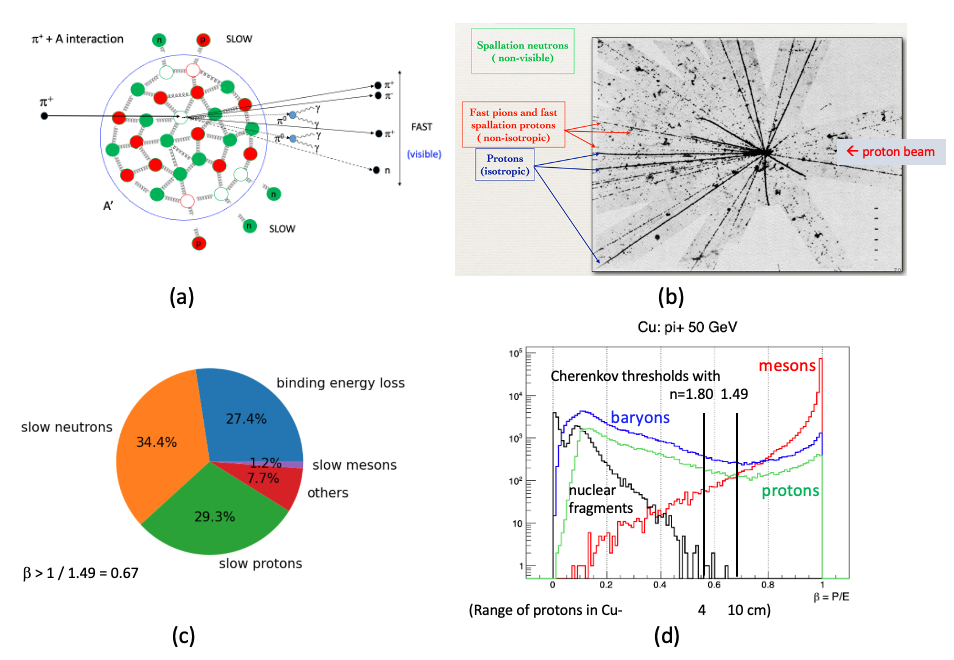}
\caption{(a) A sketch of a pion interacting with a nucleus and (b) Image of a vertex and tracks from a proton-nucleus interaction in emulsion (image from \cite{RefJ04}). 
(c) Break down of source of the invisible energy in Cherenkov calorimeter with Cu absorber for 30 GeV pion beams. The average invisible energy is 31\% of the beam energy. (d) The velocity ($\beta$) of mesons, baryons and nuclear fragments in showers in $\pi^+Cu$ interaction.  The Cherenkov thresholds for the index of refraction 1.49 (1.80) for plastic (sapphire) fibers are shown. 
}
\label{fig-1x}       
\end{figure}

A hadronic shower consists of many hadron-nucleus interactions where the interaction point can be thought of as a vertex with tracks emerging from it (Figure~\ref{fig-1x}). At each vertex, a fraction of the energy of the incoming particle is lost to break up nuclei (binding energy losses) and generate many low-energy protons and neutrons that are undetectable, stopped either in the absorber or below the Cherenkov threshold. Other sources are neutrinos, muons, baryon mass produced in baryon pair productions, and leakage of shower due to the finite size of the calorimeter. 
The ionization calorimeter detects some of those slow neutrons and protons, resulting in smaller invisible energy. 

\subsection{Vertex Counting for Invisible Energy Estimation}
\begin{figure}[ht]
\centering
\includegraphics[width=8cm,clip]{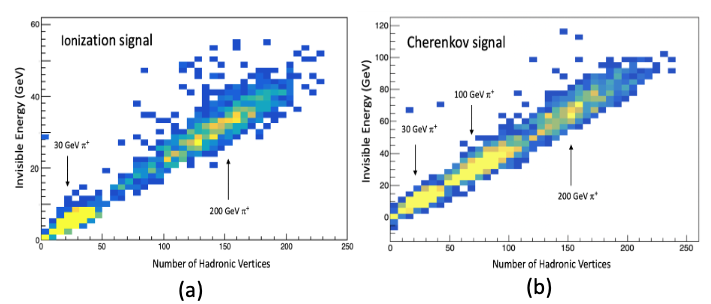}
\caption{(a) The correlation between the invisible energy and the number of hadronic vertices within the hadronic shower for 20 and 200 GeV pions where the signal is based on ionization, and (b) for 30, 100, and 200 GeV pions in the Cherenkov signal in copper absorber (figures from \cite{RefJ02}).
}
\label{fig-3x}       
\end{figure}

There are many vertices in a hadron shower and the invisible energy at each vertex varies.  It seems a daunting task to estimate the total invisible energy from these complex hadronic shower images. We simulated ionization and Cherenkov signal using GEANT4 and found a clear correlation between invisible energy and the number of inelastic hadron interaction vertices (Figure~\ref{fig-3x}). This correlation suggests a possibility of estimating the invisible energy by simply counting the number of vertices in the shower. Some AI/ML tools are known to be very efficient in analyzing ``images.'' We used Convolutional Neural Networks (CNN) and Graph Neural Networks (GNN) \cite{RefJ03} to estimate invisible energy from 3D shower images for beam energy reconstruction.
\subsection{Energy Reconstruction with GNN}
\begin{figure}[ht]
\centering
\includegraphics[width=8cm,clip]{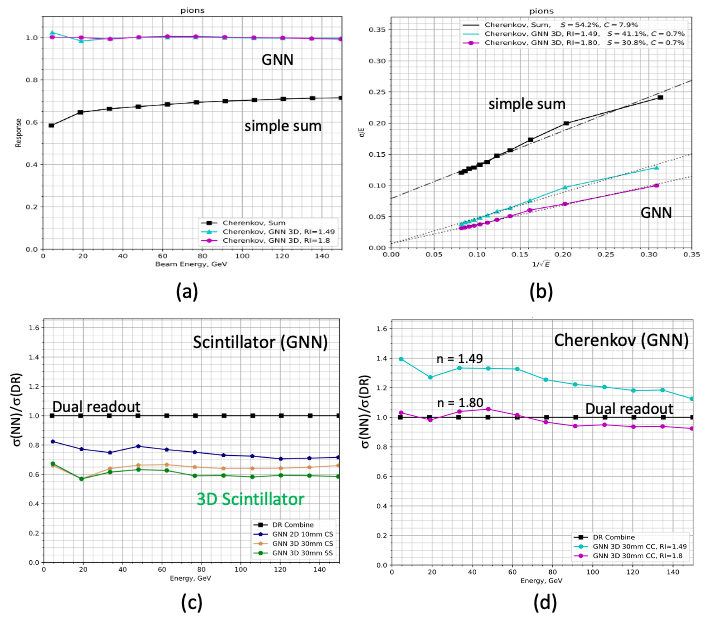}
\caption{(a) Response and (b) energy resolution of the Cherenkov fiber calorimeter for 4-150 GeV $\pi^\text{+}$ with the traditional method (simple sum) and GNN energy regression. Two cases of the index of refraction were assumed for the clear ``Cherenkov'' fiber. (c) Energy resolutions of the half-scintillator (CS) and the full scintillator (SS) calorimeters, and (d) the Cherenkov calorimeter (CC) with the GNN energy regression relative to the resolution with the Dual Readout method. The grid sizes for the GNN energy regression were 1x1 cm\textsuperscript{2} in 2D and 3x3x3 cm\textsuperscript{3} in 3D.
}
\label{fig-4x}       
\end{figure}

We simulated 3D images of the Cherenkov signal in a fiber calorimeter using GEANT4 with the FTFP\_BERT physics list. Fibers of 1 mm diameter were placed in 2 m long Cu absober with the separation of fibers of 1.5 mm (center-to-center distance). Production points of Cherenkov photons captured in fibers were recorded and fed into a GNN for energy regression.  
The energy scale was restored and the energy resolution improved by more than a factor of two (Figure~\ref{fig-4x}). This shows that we can reconstruct the hadron energy with good precision using only the fast or Cherenkov component of the shower. Furthermore, the higher index of refraction ($n$=1.80) improves the resolution compared to ($n=$1.49), indicating the importance of detecting slow isotropic protons. This observation is consistent with a result from our previous analysis with CNN Class Activation Mapping in the ionization calorimeter, which also indicated the importance of slow protons in the transverse direction around the vertex \cite{RefJ01}. 
The scintillation calorimeter detects these slow protons. We simulated scintillation fiber calorimeter in two configurations: CS (50\% Cher, 50\% scint) and SS (100\% scint). The physical layouts of the fibers were the same as the original Cherenkov fiber calorimeter, CC (100\% Cher), and the integration time was limited to 10 ns.
We observed an improvement in resolutions (Figure~\ref{fig-4x}(c)) over the Cherenkov calorimeter (Figure~\ref{fig-4x}(d)). 

\section{Choice of Hadron Calorimeter}
\label{sec-105}
\subsection{Absorber Material}
The binding energy losses in Cu are much smaller compared to that of other common absorbers ({\it e.g.} Pb, W, and U). In addition, much fewer neutrons are produced in Cu, making it a preferred absober material for imaging calorimetry.

\subsection{Sampling and Signal Integration Time}
The performances of Cu-fiber calorimeters are compared with the other types detectors in Figure 4, and we draw the following conclusions so far from our work: $1$) it is clear that fine sampling is key to good energy resolution as shown in Figure 4; $2$) the short integration time is desirable for the mitigation of beam-induced-background at a future Muon Collider and pile-up effects at high rate hadron colliders, $3$) the experience from the CMS forward calorimeter shows that Cherenkov calorimeter with quartz fibers is radiation-hard, and $4$) as we argue, the use of proper neural networks results in good energy resolution. The scintillation fiber calorimeter with a similarly tuned neural network, where the contribution of the slow protons to the signal is sizable, provides the best energy resolution; see Figure 4.

\begin{figure}
\centering
\includegraphics[width=7cm,clip]{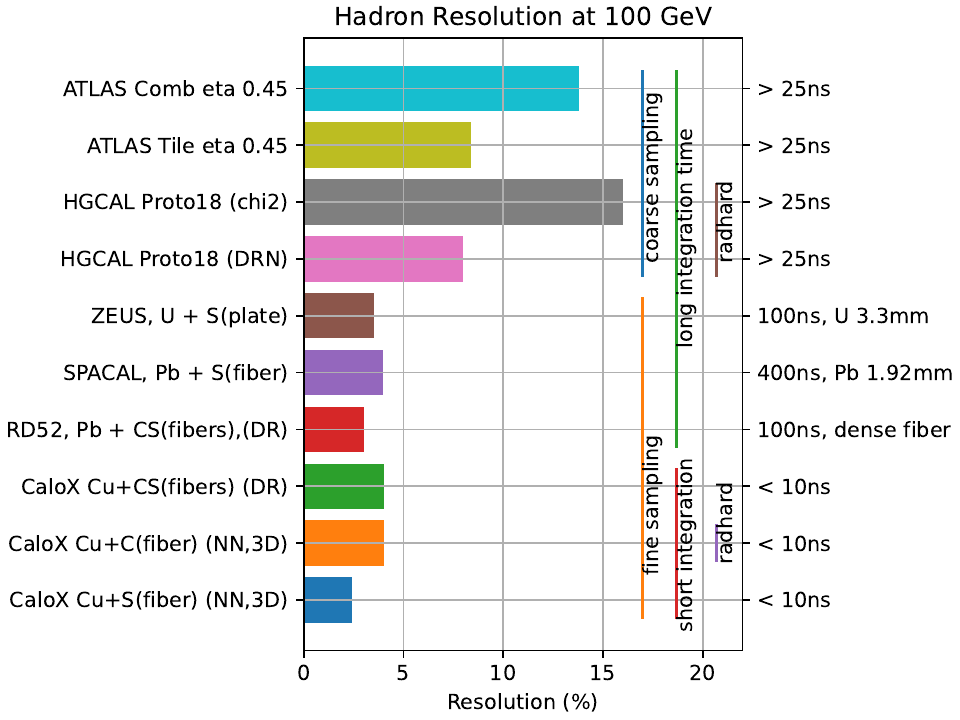}
\caption{
The compilation of hadronic energy resolution performance of various operating, planned, or simulated calorimeters at 100 GeV 
(\cite{RefJ05,RefJ06,RefJ07,RefJ08,RefJ09}). Note that the coarse sampling (every O(cm) of absorber), and the longer integration-time (>10 ns) calorimeters are grouped in the upper part of the plot, and the fine sampling (every O(mm) of absorber) and the shorter integration-time (<10 ns) are grouped in the lower part of the plot. }
\label{fig-110}       
\end{figure}

\subsection{Homogeneous Calorimeter}
\begin{figure}
\centering
\includegraphics[width=8cm,clip]{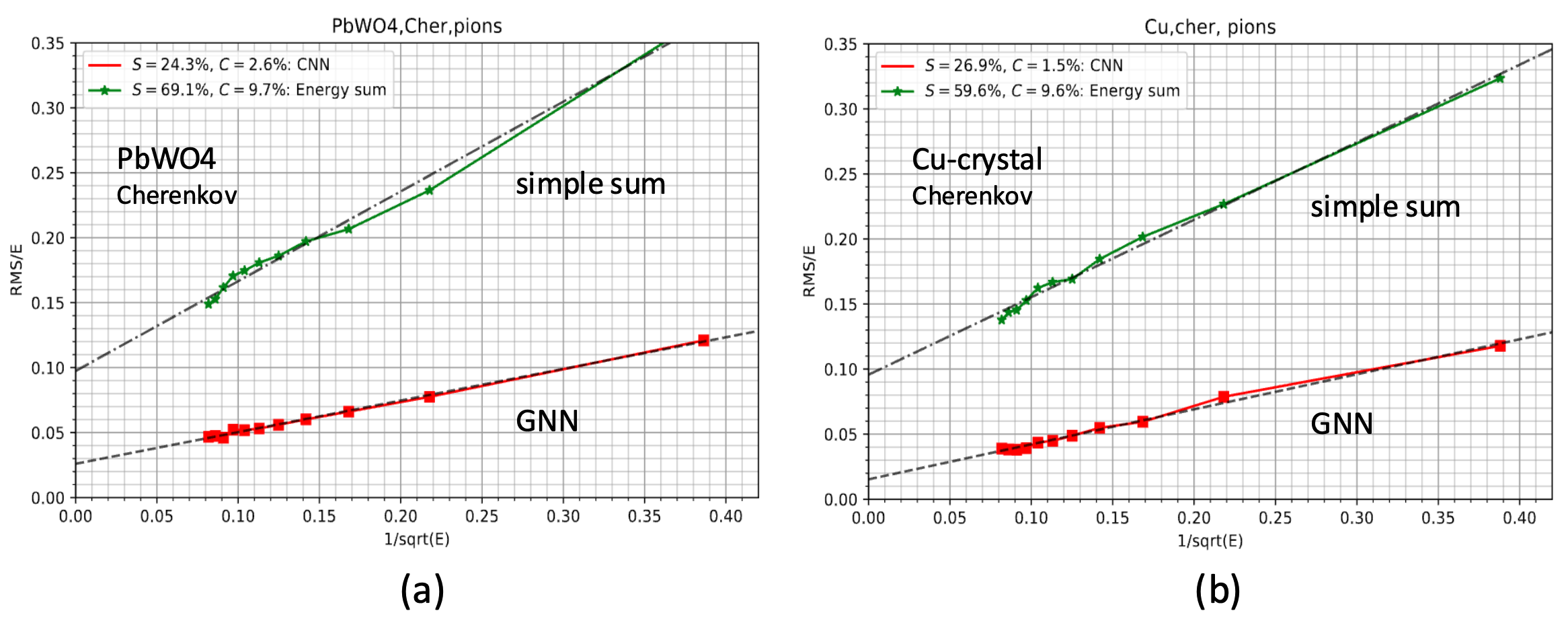}
\caption{The resolution of of the Cherenkov crystal calorimeter for 4-150 GeV $\pi^+$. (a) PbWO4 ($n=$2.2) and (b) "Cu crystal". A Cu block is treated as an imaginary crystal of $n=$1.49.}
\label{fig-xtal}       
\end{figure}

Homogeneous calorimeter is an ``ultimate sampling calorimeter.''
We simulated Cherenkov signal in a block of PbWO4 crystal ($n=$2.2) and a Cu block, which was treated as an imaginary crystal of $n=$1.49.
The energy resolution of Cu with CNN is better than that of PbWO4. (Figure~\ref{fig-xtal}).
Segmented “Crystal” of 2$\times$2$\times$2 cm$^3$ does not provide very significant improvement in resolution over the
densely packed fiber calorimeter. This indicates that the resolution seen in the fiber calorimeter case was not limited by the sampling but is limited by the method of the invisible energy estimation with CNN/GNN. Further analysis of the effectiveness of CNN/GNN may provide clues to improve the resolution of the fiber calorimeter and reveal the limits of homogeneous hadron calorimetry. 

\section{Potential of Cherenkov Fiber Calorimeter}
\subsection{HG-DREAM}
The original DREAM module was constructed in 2002-2003 at TTU and used at CERN to test the DR calorimetry in 2004-2012. It was brought back to TTU in 2023 and will be used as a test bed for new technologies and new ideas. The module has been refurbished with SiPMs for highly granular readout and data taking with cosmic muons has started. The module will be placed on test beams to verify the vertex imaging calorimetry concept and its further development. See Figure~\ref{fig-hgdream}a for fiber readout configurations and \cite{RefJ10} for more information on HG-DREAM.

\subsection{Dual Cherenkov Fiber Readout}
\begin{figure}
\centering
\includegraphics[width=8cm,clip]{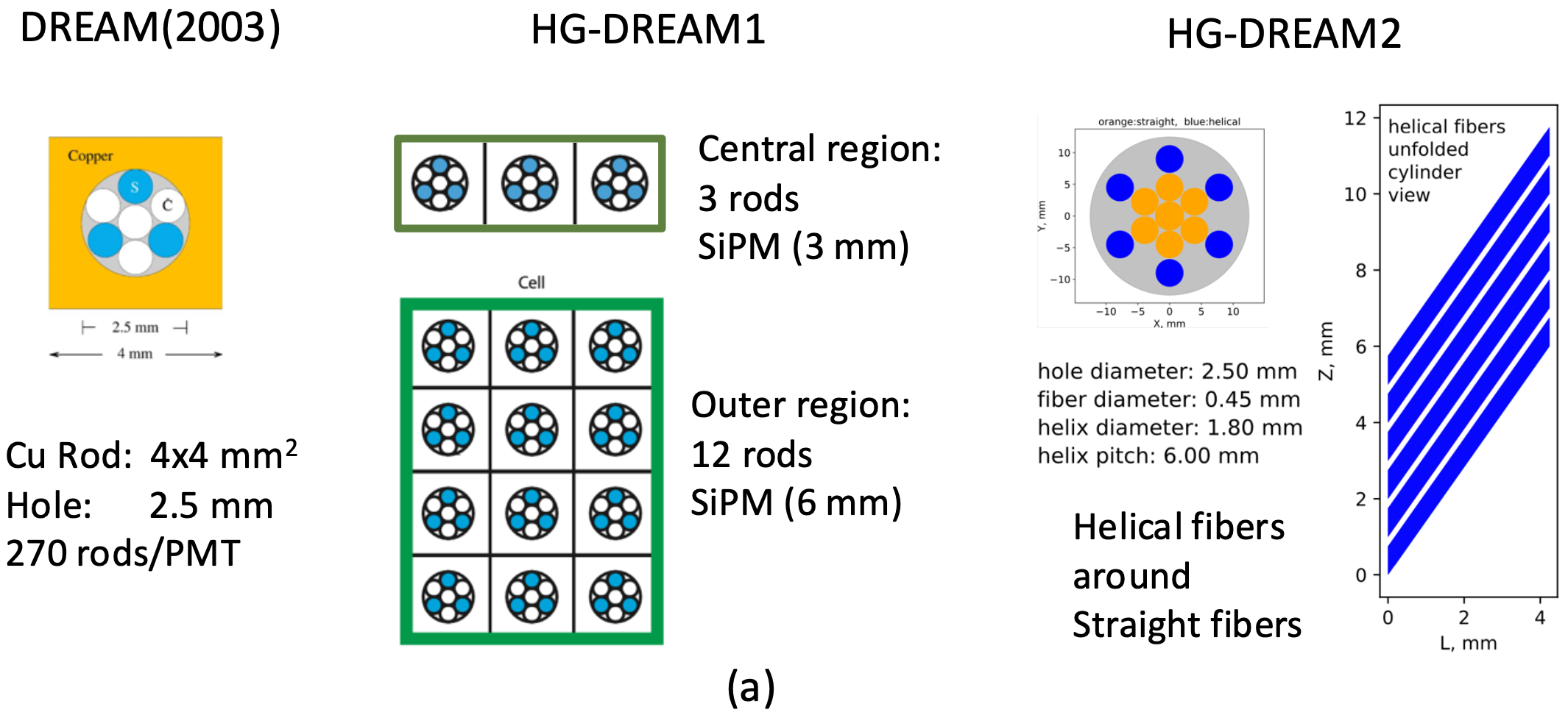}
\includegraphics[width=8cm,clip]{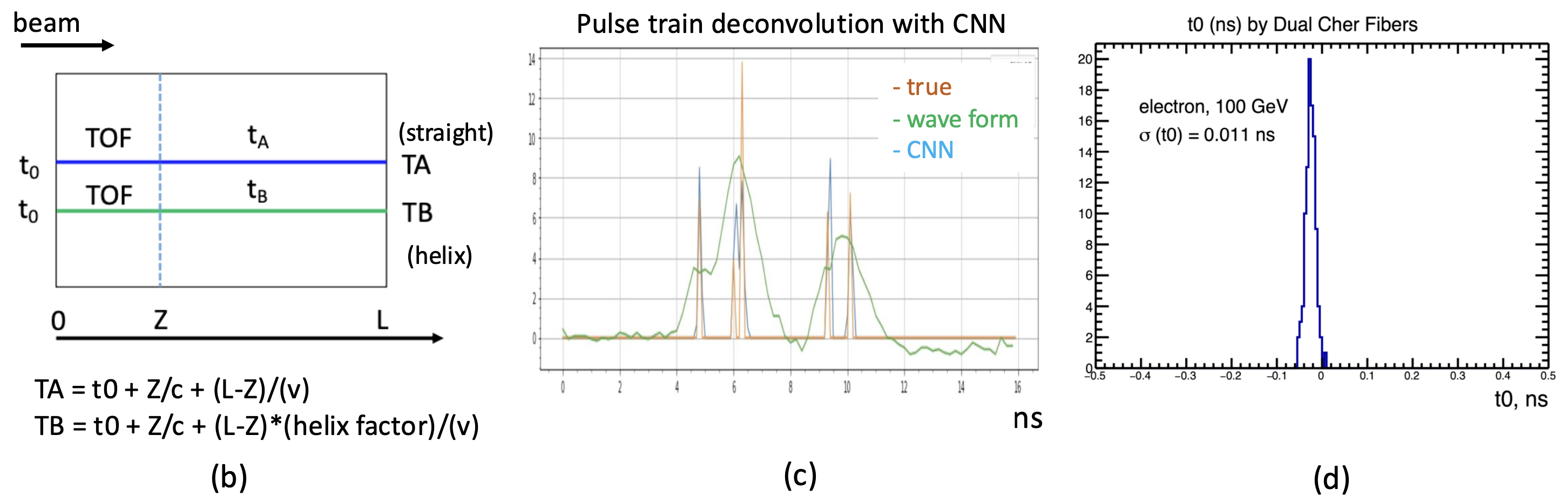}
\caption{(a) The DREAM module consists of $\sim$5,000 Cu rods of 2-m long and 4$\times$4 mm$^2$ cross section with a 2.5-mm diameter hole. Each rod contains 4 Cherenkov fibers and 3 scintillation fibers in the hole. The original DREAM used coarse readout: 270 rods per tower ($\sim$3.7 cm in radius) with PMT readout. The HG-DREAM uses high granularity readout: 3x1 (4x3) rods per SiPM readout at the center (outer) region of the module. HG-DREAM2 tests the Dual Cherenkov Fiber Readout (DCFR) with helical fibers around straight fibers in the rod. (b) concept of TOF and Z position measurement in DCFR. (c) Reconstruction of the timing of multi-hits with CNN. (d) Reconstructed t0 with GNN.}
\label{fig-hgdream}       
\end{figure}

Fiber calorimeter is a 2D ($x,y$) segmented calorimeter. The arrival time of the signal at SiPM gives the third coordinate ($z$) (see the equation in Figure~\ref{fig-hgdream}b). CNN can be used to resolve the timing of multi hits along the fibers (Figure~\ref{fig-hgdream}c). 
The original scheme of ``Longitudinal Segmentation with Timing'' assumes $t0$ and $v$ are known, {\it i.e.} $t0$ from the beam collision time of the accelerator and $v=c/n$, where $c$ is the speed of light in vacuum and $n$ the index of refraction of the fiber (1.49 for plastic fiber). With a pair of fibers with different velocities $VA$ and $VB$, we can measure both $t0$ and $z$. We simulated helix fibers around straight fibers. Since the path length of light in the helix is longer, the effective velocity along the $z$-axis is slower, $VB(z)=VA(z)/1.38$ in the simulated case, where 1.38 is the ``helix factor'' corresponding to the path length difference. The reconstructed $t0$ is shown in Figure~\ref{fig-hgdream}d. This suggests a possibility to build a calorimeter with Dual Cherenkov Fiber Readout (DCFR) for the TOF measurement with resolution up to 10 ps in its full body. Instead of helix fiber, fibers with a higher index of refraction, for example Sapphire, $n=$1.78, may be used in the DCFR calorimeter.

\subsection{3D Local Clustering with AI/ML}
\begin{figure}
\centering
\vspace*{1cm}       
\includegraphics[width=8cm,clip]{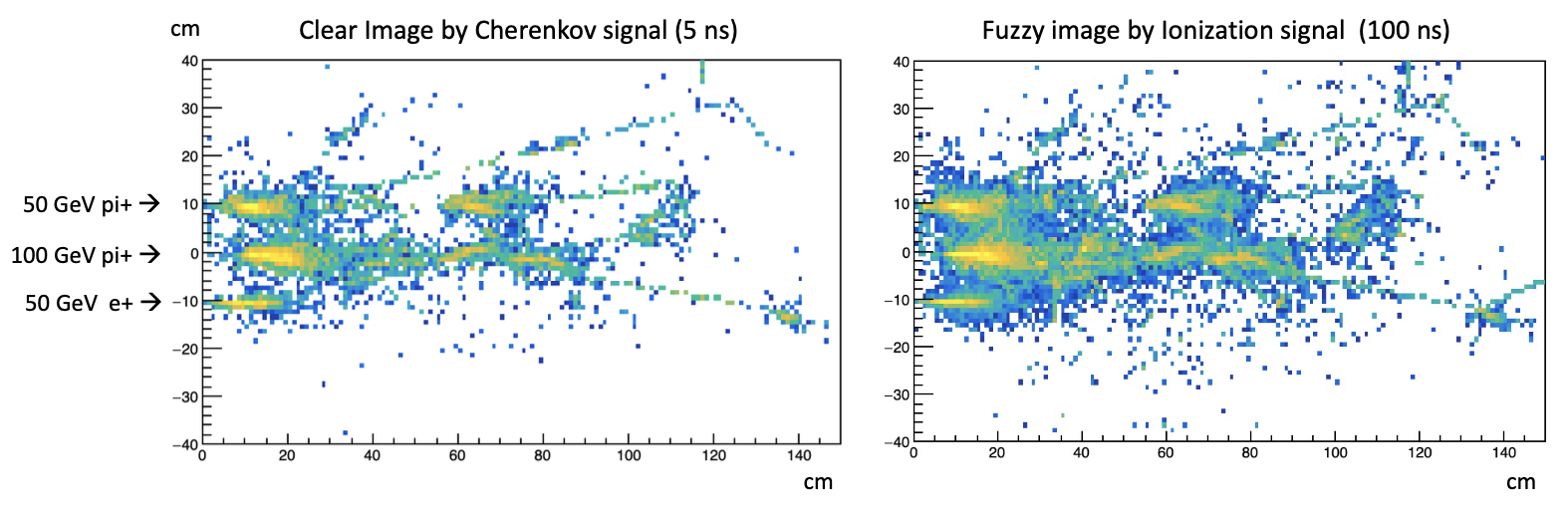}
\caption{Example of shower images of a ``pseudo jet.'' Three particles hit the calorimeter 10 cm apart. The image with the Cherenkov signal (left) is cleaner than with ionization signal (right).  Assignment of the cluster of hits to each incident particle can be performed more effectively with the clean Cherenkov signal rather than the ionization signal.}
\label{fig-6x}       
\end{figure}

The Cherenkov signal produces a clear and narrow image of the hadronic shower compared to the ionization signal (Figure~\ref{fig-6x}). The AI/ML approaches may use these cleaner shower images for 3D pattern recognition of complex event structures, for example: 
\begin{itemize}
\item Reconstruction of individual particles in jet (then link to inner tracks for the particle flow reconstruction).
\item analysis of jet substructure to identify boosted $W/ Z$ / top / Higgs in multi-TeV collider experiments.
\end{itemize}

Furthermore, it is conceivable to build an AI-ASIC system on the detector to perform a series of event-reconstruction tasks using very localized data. If such a system is realized, the transmission of data from the front-end to the back-end will be greatly reduced.

\section{Conclusions}
\label{sec-140}
Vertex imaging calorimetry using AI / ML may provide very fast hadron calorimetry in future collider experiments. The integration time of the signal will be below 10 ns.  The
Dual Cherenkov Fiber Readout has good potential to build a 4D ($xyzt;E$) imaging calorimeter with a time resolution of O(10 ps) in its full body.  So far we have used simplistic CNN and GNN algorithms. The more optimized use of AI/ML will bring further improvements to future calorimetry.

\section{Acknowledgement}
We are grateful for the financial support from the US Department of Energy, Office of Science grants DE-SC0015592 and DE-SC0023690, and Texas Tech University.

%
%
%

\end{document}